\documentclass{llncs}
\usepackage{llncsdoc}

\usepackage{tabularx}

\usepackage{times}
\usepackage{graphicx}
\usepackage{array}
\usepackage{amsmath}
\usepackage{amssymb}
\usepackage{algorithm}
\usepackage{algorithmic}
\usepackage{url}
\usepackage{multirow}
\usepackage{subfigure}

\usepackage[export]{adjustbox}

\newcolumntype{V}{>{$\vcenter\bgroup\hbox\bgroup}c<{\egroup\egroup$}}
\def\Hline{\noalign{\hrule height 4\arrayrulewidth}}

\graphicspath{{./imgs/}}

\begin{document}

\title{Stream Quantiles via Maximal Entropy Histograms}
\author{Ognjen~Arandjelovi\'c, Ducson~Pham, and~Svetha~Venkatesh}
\institute{Centre for Pattern Recognition and Data Analytics, Deakin University}

\maketitle

\begin{abstract}
We address the problem of estimating the running quantile of a data stream when the memory for storing observations is limited. We (i) highlight the limitations of approaches previously described in the literature which make them unsuitable for non-stationary streams, (ii) describe a novel principle for the utilization of the available storage space, and (iii) introduce two novel algorithms which exploit the proposed principle. Experiments on three large real-world data sets demonstrate that the proposed methods vastly outperform the existing alternatives.
\end{abstract}

\section{Introduction}
The problem of quantile estimation is of pervasive importance across a variety of signal processing applications. It is used in data mining, simulation modelling~\cite{JainChla1985}, database maintenance, risk management in finance~\cite{SgouYaoYast2013}, and understanding computer network latencies~\cite{BuraSuri2009}, amongst others. A particularly challenging form of the quantile estimation problem arises when the desired quantile is high-valued (i.e.\ close to 1, corresponding to the tail of the underlaying distribution) and when data needs to be processed as a stream, with limited memory capacity. An illustrative practical example of when this is the case is encountered in CCTV-based surveillance systems. In summary, as various types of low-level observations related to events in the scene of interest arrive in real-time, quantiles of the corresponding statistics for time windows of different durations are needed in order to distinguish `normal' (common) events from those which are in some sense unusual and thus require human attention. The amount of incoming data is extraordinarily large and the capabilities of the available hardware highly limited both in terms of storage capacity and processing power.

\subsection{Previous work}\label{ss:prev}
Unsurprisingly, the problem of estimating a quantile of a set has received considerable research attention, much of it in the realm of theoretical research. In particular, a substantial amount of work has focused on the study of asymptotic limits of computational complexity of quantile estimation algorithms~\cite{GuhaMcGr2009,MunrPate1980}. An important result emerging from this corpus of work is the proof by Munro and Paterson~\cite{MunrPate1980} which in summary states that the working memory requirement of any algorithm that determines the median of a set by making at most $p$ sequential passes through the input is $\Omega(n^{1/p})$ (i.e.\ asymptotically growing at least as fast as $n^{1/p}$). This implies that the exact computation of a quantile requires $\Omega(n)$ working memory. Therefore a single-pass algorithm, required to process streaming data, will necessarily produce an estimate and not be able to guarantee the exactness of its result.

Most of the quantile estimation algorithms developed for the use in practice are not single-pass algorithms i.e.\ cannot be applied to streaming data \cite{GuraSriv1990}. On the other hand, many single-pass approaches focus on the exact computation of the quantile and thus demand $O(n)$ storage space which is clearly an unfeasible proposition in the context we consider in the present paper. Amongst the few methods described in the literature and which satisfy our constraints are the histogram-based method of Schmeiser and Deutsch~\cite{SchmDeut1977} (also by McDermott \textit{et al.} \cite{McDeBabuLiecLin2007}), and the $P^2$ algorithm of Jain and Chlamtac~\cite{JainChla1985}. Schmeiser and Deutsch maintain a preset number of bins, scaling their boundaries to cover the entire data range as needed and keeping them equidistant. Jain and Chlamtac attempt to maintain a small set of \textit{ad hoc} selected key points of the data distribution, updating their values using quadratic interpolation as new data arrives. Lastly, random sample methods, such as that described by Vitter~\cite{Vitt1985}, and Cormode and Muthukrishnan~\cite{CormMuth2005}, use different sampling strategies to fill the available buffer with random data points from the stream, and estimate the quantile using the distribution of values in the buffer.

In addition to the \textit{ad hoc} elements of the previous algorithms for quantile estimation on streaming data, which itself is a sufficient cause for concern when the algorithms need to be deployed in applications which demand high robustness and well understood failure modes, it is also important to recognize that an implicit assumption underlying these approaches is that the data is governed by a stationary stochastic process. The assumption is often invalidated in real-world applications. As we will demonstrate in Sec.~\ref{s:eval}, a consequence of this discrepancy between the model underlying existing algorithms and the nature of data in practice is a major deterioration in the quality of quantile estimates. Our principal aim is thus to formulate a method which can cope with non-stationary streaming data in a more robust manner.

\section{Proposed algorithms}
We start this section by formalizing the notion of a quantile. This is then followed by the introduction of the key premise of our contribution and finally a description of two algorithms which exploit the underlying idea in different ways. The algorithms are evaluated on real-world data in the next section.

\subsection{Quantiles}
Let $p$ be the probability density function of a real-valued random variable $X$. Then the $q$-quantile $x_q$ of $p$ is defined as:
{\small\begin{align}
  \int_{-\infty}^{x_q} p(x)~dx = q.
\end{align}}
Similarly, the $q$-quantile of a finite set $D$ can be defined as:
{\small\begin{align}
  \left|\{ x~:~ x \in D \text{ and  } x \leq x_p\}\right| \leq q \times |D|.
\end{align}}
In other words, the $q$-quantile is the smallest value below which $q$ fraction of the total values in a set lie.

\subsection{Maximal entropy histograms}\label{ss:maxEnt}
A consequence of the non-stationarity of data streams that we are dealing with is that at no point in time can it be assumed that the historical distribution of data values is representative of its future distribution, regardless of how much data has been seen. Thus, the value of a particular quantile can change greatly and rapidly, in either direction (i.e.\ increase or decrease). To be able to adapt to such unpredictable variability in input it is therefore not possible to focus on only a part of the historical data distribution but rather it is necessary to store a `snapshot' of the entire distribution. We achieve this using a histogram of a fixed length, determined by the available working memory. In contrast to the previous work which either distributes the bin boundaries equidistantly or uses \textit{ad-hoc} adjustments, our idea is to maintain bins in a manner which maximizes the entropy of the corresponding estimate of the historical data distribution.

\subsection{Method 1: interpolated bins}\label{ss:method1}
The first method we describe readjusts the boundaries of a fixed number of bins after the arrival of each new data point $d_{i+1}$. Without loss of generality let us assume that each each datum is positive i.e.\ that $d_i > 0$. Furthermore, let the upper bin boundaries before the arrival of $d_i$ be $b^i_1, b^i_2,\ldots, b^i_n$, where $n$ is the number of available bins. Thus, the $j$-th bin's catchment range is $(b^i_{j-1}, b^i_j]$ where we will take that $b^i_0=0$ for all $i$. We wish to maintain the condition that the piece-wise uniform probability density function approximation of the historical data distribution described by this histogram has the maximal entropy of all those possible with the histogram of the same length. This is achieved by having equiprobable bins. Thus, before the arrival of $d_{i+1}$, the number of historical data points in each bin is the same and equal to $i/n$. The corresponding cumulative density is given by:
{\small\begin{align}
  p^i(d)=\frac{1}{n} \times \left[j + \frac{d-b^i_{j-1}}{b^i_j-b^i_{j-1}} \right]  \text{~~~~~and~~~~~} b^i_{j-1} < d \leq b^i_j.
\end{align}}

After the arrival of $d_i$ but before the readjustment of bin boundaries, the cumulative density becomes:
{\small\begin{align}
  \widetilde{p}^i(d)= \begin{cases}
            \frac{i}{i+1} \times \frac{1}{n} \times \left[j + \frac{d-b^i_{j-1}}{b^i_j-b^i_{j-1}} \right] ~~~~~~~~~~~~~~\text{ for } d < d_i\\
            \frac{i}{i+1} \times \frac{1}{n} \times \left[j + \frac{d-b^i_{j-1}}{b^i_j-b^i_{j-1}} \right] +\frac{1}{i+1} ~~~~\text{ for } d \geq d_i\\
          \end{cases}
\end{align}}
Lastly, to maintain the invariant of equiprobable bins, the bin boundaries are readjusted by linear interpolation of the corresponding inverse distribution function.

\subsection{Method 2: data-aligned bins}\label{ss:method2}
The algorithm described in the proceeding section appears optimal in that it always attempts to realign bins so as to maintain maximal entropy of the corresponding approximation for the given size of the histogram. However, a potential source of errors can emerge cumulatively as a consequence of repeated interpolation, done after every new datum. Indeed, we will show this to be the case empirically in Sec.~\ref{s:eval}. We now introduce an alternative approach which aims to strike a balance between some unavoidable loss of information, inherently a consequence of the need to readjust an approximation of the distribution of a continually growing data set, and the desire to maximize the entropy of this approximation.

Much like in the previous section, bin boundaries are potentially altered each time a new datum arrives. There are two main differences in how this is performed. Firstly, unlike in the previous case, bin boundaries are not allowed to assume arbitrary values; rather, they are constrained to the values of the seen data points. Secondly, only at most a single boundary is adjusted for each new datum. We now explain this process in detail.

\begin{figure}[htb]
  \centering
  \subfigure[]{\includegraphics[width=0.36\textwidth]{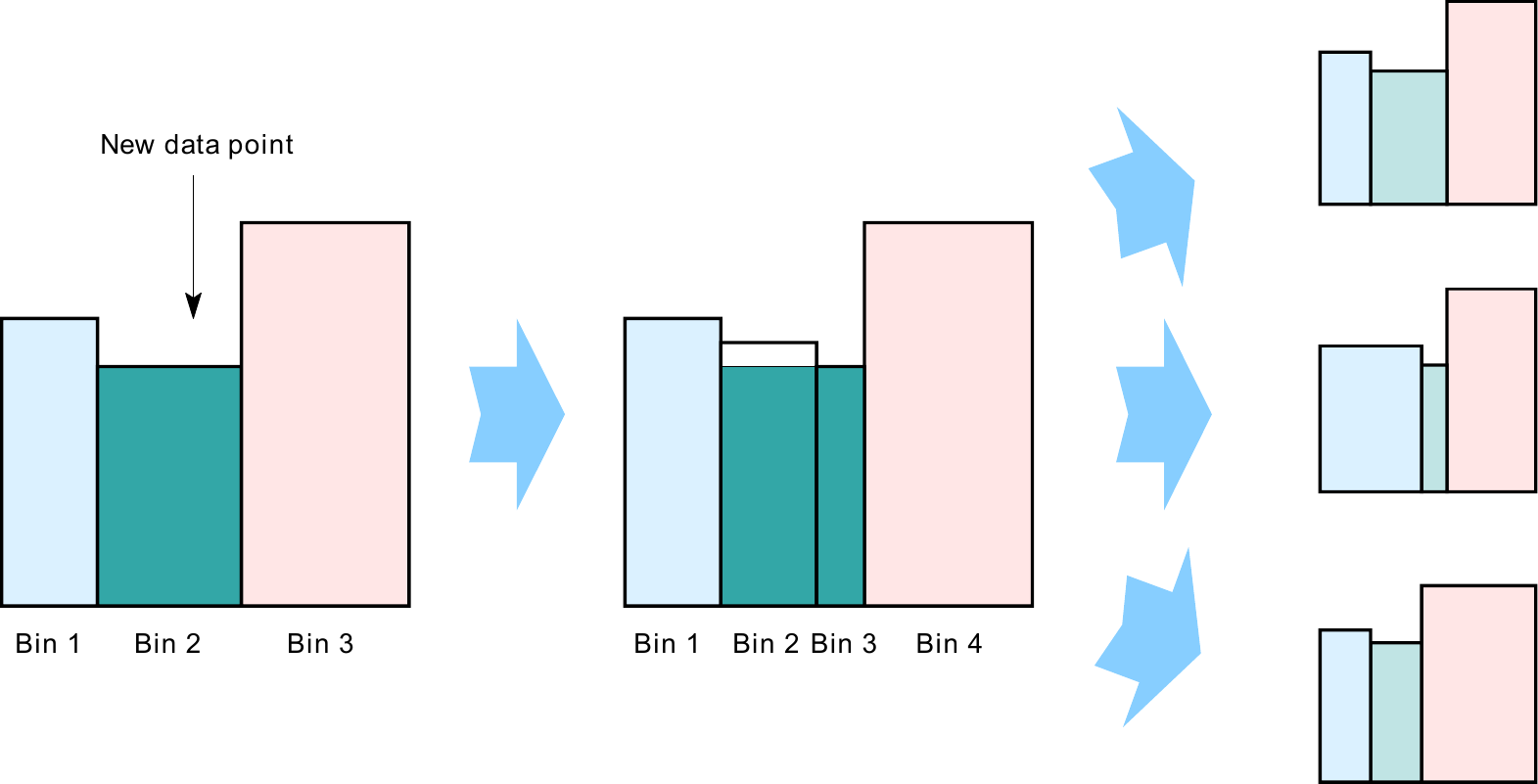}\label{f:merge}}~~~~~~-
  \subfigure[]{\includegraphics[width=0.55\textwidth]{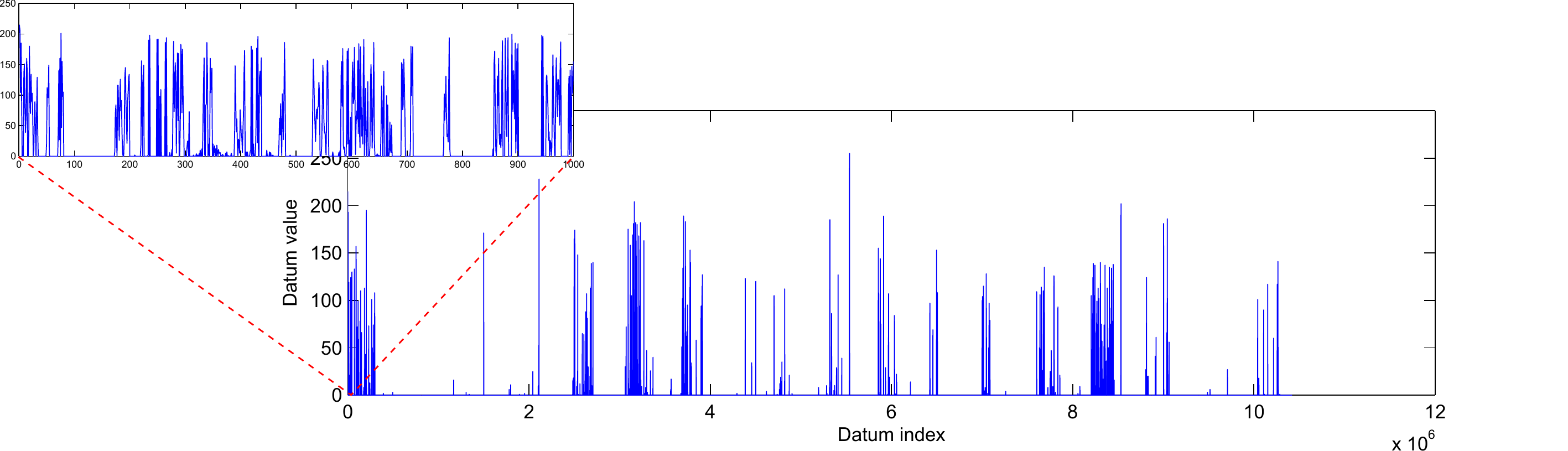}\label{f:streams}}
  \vspace{-5pt}
  \caption{ (a) The update step in our data-aligned adaptive histogram algorithm. (b) One of the three large data streams used in our evaluation. }
  \vspace{-10pt}
\end{figure}

As before, let the upper bin boundaries before the arrival of a new data point be $b^i_1, b^i_2,\ldots, b^i_n$. Since unlike in the case of the previous algorithm in general the bins will not be equiprobable we also have to maintain a corresponding list $c^i_1, c^i_2,\ldots, c^i_n$ which specifies the corresponding data counts. Each time a new data point arrives a new, an $(n+1)$-st bin is created temporarily. If the value of the new datum is greater than $b^i_n$ (and thus greater than any of the historical data), a new bin is created after the current $n$-th bin, with the upper boundary set at $d(i)$. The corresponding datum count $c$ of the bin is set to 1. Alternatively, if the value of the new data point is lower than $b^i_n$ then there exists $j$ such that $b^i_{j-1} < d \leq b^i_j$ and the new bin is inserted between the $(j-1)$-st and $j$-th bin. Its datum count is estimated as follows:
{\small\begin{align}
  c = c_j \times \frac{d-b^i_{j-1}}{b^i_j-b^i_{j-1}} + 1.
\end{align}}
Thus, regardless of the value of the new data point, temporarily the number of bins is increased by 1. The original number of bins is then restored by merging exactly a single pair of neighbouring bins. For example, if the $k$-th and $(k+1)$-st bin are merged, the new bin has the upper boundary value set to the upper boundary value of the former $(k+1)$-st bin, i.e.\ $b^i_{k+1}$, and its datum count becomes the sum of counts for the $k$-th and $(k+1)$-st bins, i.e.\ $c^i_k + c^i_{k+1}$.  The choice of which neighbouring pair to merge, out of $n$ possible options, is made according to the principle stated in Sec.~\ref{ss:maxEnt}, i.e.\ the merge actually performed should maximize the entropy of the new $n$-bin histogram. This is illustrated conceptually in Fig.~\ref{f:merge}.

\section{Evaluation and results}\label{s:eval}
To assess the effectiveness of the proposed algorithms, we evaluated their performance on three large `real-world' data streams. The streams correspond to motion statistics used by an existing CCTV surveillance system for the detection of abnormalities in video footage. It is important to emphasize that the data we used was not acquired for the purpose of the present work nor were the cameras installed with the same intention. Rather, we used data which was acquired using existing, operational surveillance systems. In particular, our data comes from three CCTV cameras, two of which are located in Mexico and one in Australia. We next explain the source of these streams and the nature of the phenomena they represent.

\subsection{Real-world surveillance data}
Computer-assisted video surveillance data analysis is of major commercial and law enforcement interest. On a broad scale, systems currently available on the market can be grouped into two categories in terms of their approach. The first group focuses on a relatively small, predefined and well understood subset of events or behaviours of interest such as the detection of unattended baggage, violent behaviour, etc \cite{Phil,LaveKhanThur2007}. The narrow focus of these systems prohibits their applicability in less constrained environments in which a more general capability is required. In addition, these approaches tend to be computationally expensive and error prone, often requiring fine tuning by skilled technicians. This is not practical in many circumstances, for example when hundreds of cameras need to be deployed as often the case with CCTV systems operated by municipal authorities. The second group of systems approaches the problem of detecting suspicious events at a semantically lower level~\cite{Aran2011a,Inte,iCet}. Their central paradigm is that an unusual behaviour at a high semantic level will be associated with statistically unusual patterns (also `behaviour' in a sense) at a low semantic level -- the level of elementary image/video features. Thus methods of this group detect events of interest by learning the scope of normal variability of low-level patterns and alerting to anything that does not conform to this model of what is expected in a scene, without `understanding' or interpreting the nature of the event itself. These methods uniformly start with the same procedure for feature extraction. As video data is acquired, firstly a dense optical flow field is computed. Then, to reduce the amount of data that needs to be processed, stored, or transmitted, a thresholding operation is performed. This results in a sparse optical flow field whereby only those flow vectors whose magnitude exceeds a certain value are retained; non-maximum suppression is applied here as well. Normal variability within a scene and subsequent novelty detection are achieved using various statistics computed over this data.  A typical data stream, shown partially in Fig.~\ref{f:streams}, corresponds to the values of such a statistic. Observe the non-stationary nature of the data streams which is evident both on the long and short time scales.

\subsection{Results}
\begin{figure}[htb]
  \centering
  \subfigure[Proposed 1]{\includegraphics[trim=0cm 0cm 22cm 0cm, clip=true, width=0.18\textwidth]{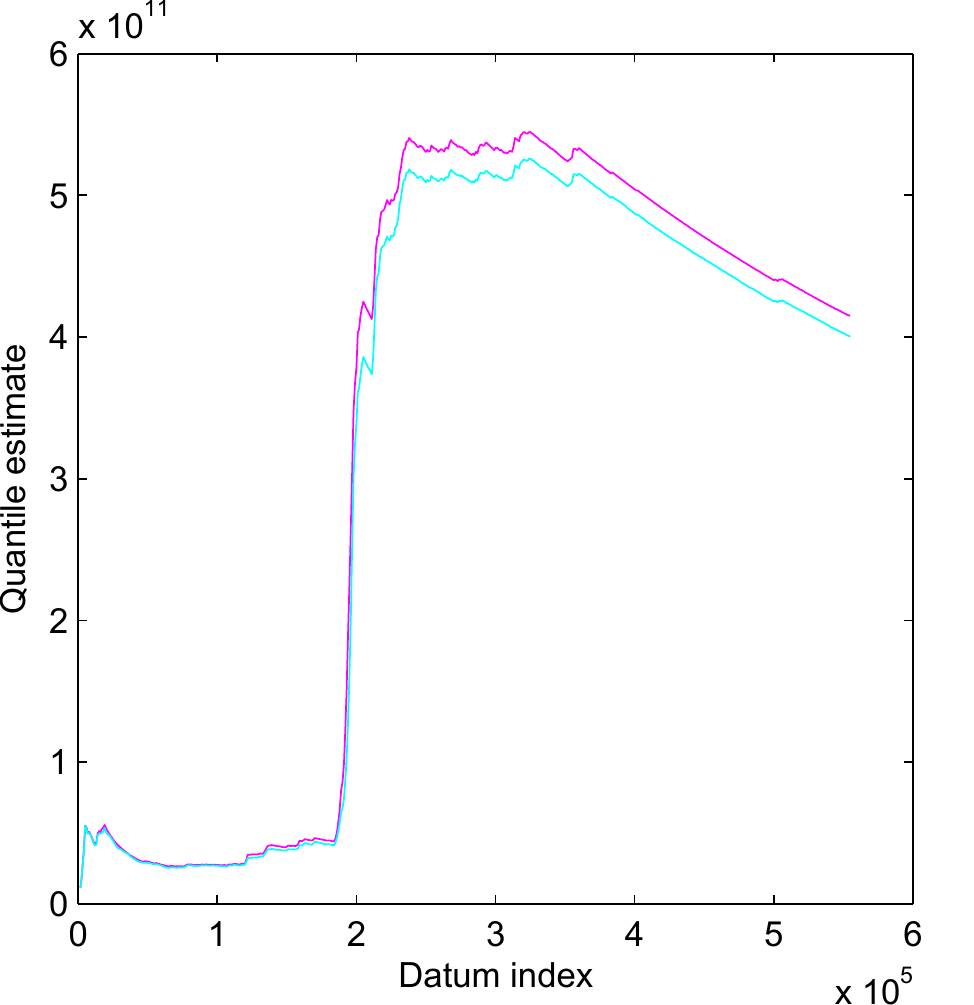}}~
  \subfigure[Proposed 2]{\includegraphics[trim=0cm 0cm 22cm 0cm, clip=true, width=0.18\textwidth]{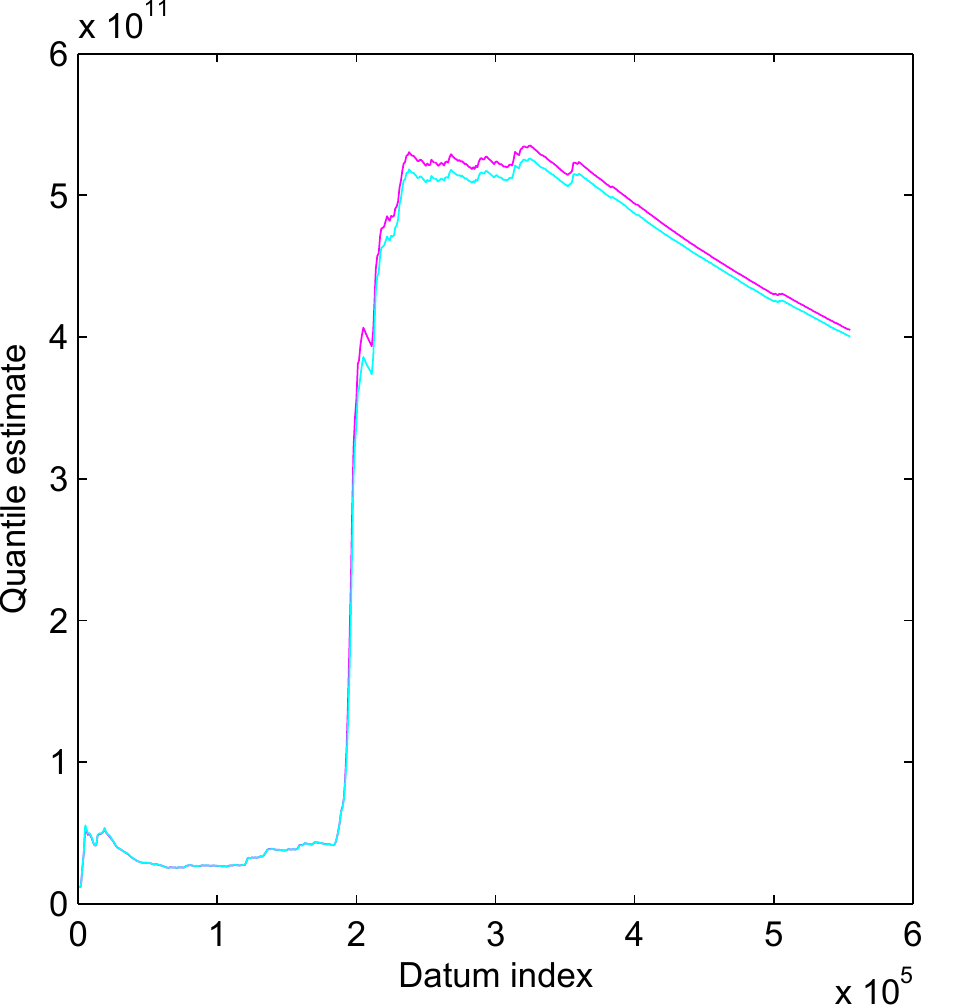}}~
  \subfigure[$P^2$ algorithm]{\includegraphics[trim=0cm 0cm 22cm 0cm, clip=true, width=0.18\textwidth]{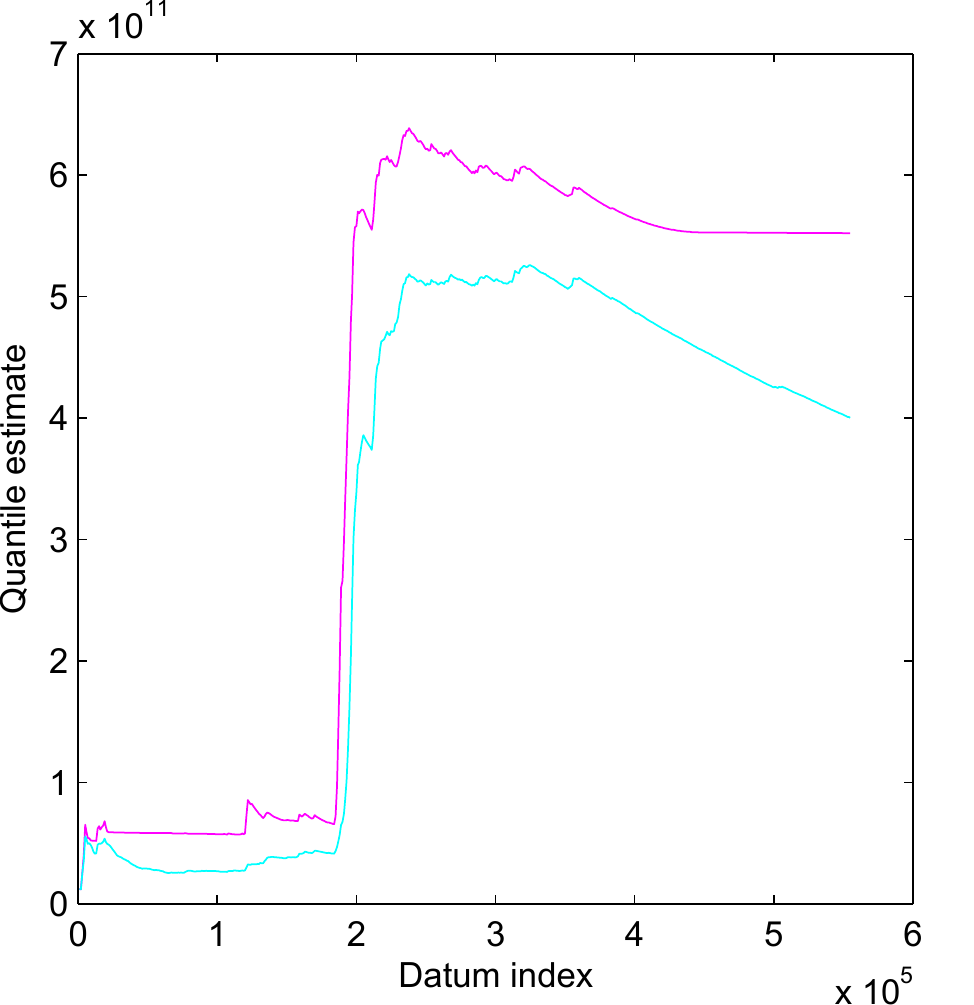}}~
  \subfigure[Rnd sample]{\includegraphics[trim=0cm 0cm 22cm 0cm, clip=true, width=0.18\textwidth]{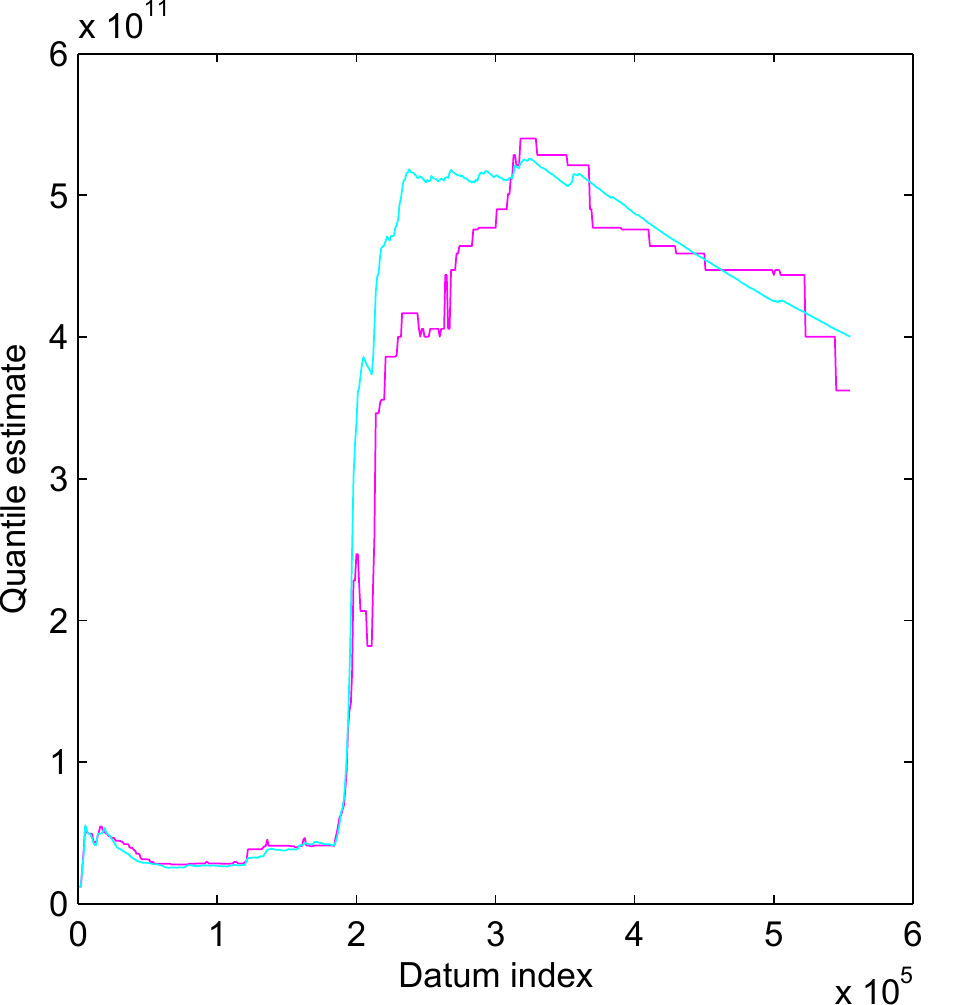}}~
  \subfigure[Uniform hist]{\includegraphics[trim=0cm 0cm 22cm 0cm, clip=true, width=0.185\textwidth]{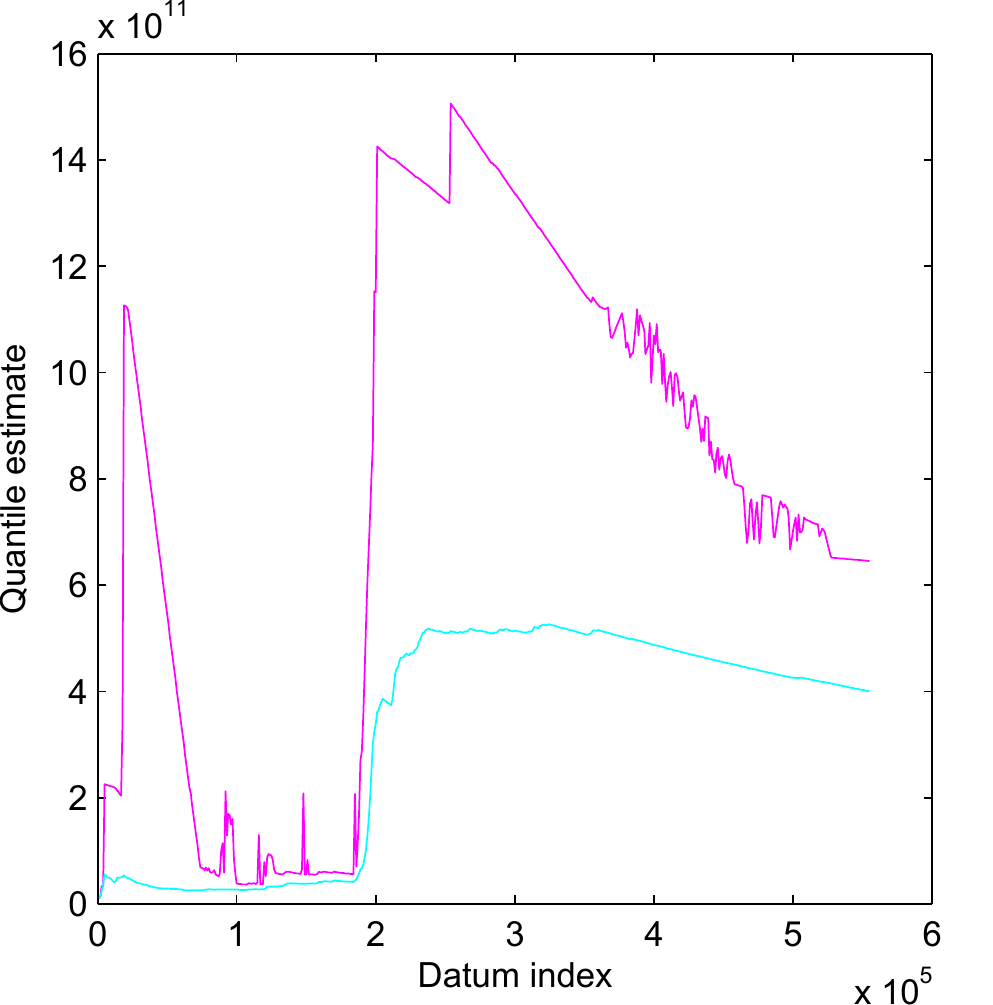}}
  \vspace{-5pt}
  \caption{ Running estimate (purple) of the 0.95-quantile on stream~1 (ground truth is shown in cyan). }
  \label{f:res095_x1}
  \vspace{-10pt}
\end{figure}

We started by comparing the performance of our algorithms with the three alternatives from the literature described in Sec.~\ref{ss:prev}: (i) the $P^2$ algorithm of Jain and Chlamtac~\cite{JainChla1985}, (ii) the random sample based algorithm of Vitter~\cite{Vitt1985}, and (iii) the uniform adjustable histogram of Schmeiser and Deutsch~\cite{SchmDeut1977}. Representative results, obtained using the same number of bins $n=500$, for 0.95-quantile on stream~1 are shown Fig.~\ref{f:res095_x1} -- the running quantile estimate of the algorithm (purple) is superimposed to the ground truth (cyan). Firstly, compare the performances of the two proposed algorithms. We found that in all cases and across time, the data-aligned bins algorithm produced a more reliable estimate. Thus, the argument put forward in Sec.~\ref{ss:method2} turned out to be correct -- despite the attempt of the interpolated bins algorithm to maintain exactly a maximal entropy approximation to the historical data distribution, the advantages of this approach are outweighed by the accumulation of errors caused by repeated interpolations. The data-aligned algorithm consistently exhibited outstanding performance on all three data sets, its estimate being virtually indistinguishable from the ground truth. This is witnessed and more easily appreciated by examining the plots showing its running relative error. In most cases the error was approximately 0.2\%; the only instance when the error would exceed this substantially is transiently at times of sudden large change in the quantile value (as in the case of stream~1), quickly recovering thereafter.

All of the algorithms from the literature performed significantly worse than both of the proposed methods. The limitations of the assumption of stationary data statistics implicitly made in the $P^2$ algorithm and discussed in Sec.~\ref{ss:prev} is readily evident by its observed performance. Following the initially good estimates when the true quantile value is relatively large, the algorithm is unable to adjust sufficiently to the changed data distribution and the decreasing quantile value. Across the three data sets, the random sample algorithm of Vitter~\cite{Vitt1985} overall performed best of the existing methods, never producing a grossly inaccurate estimate. Nonetheless its accuracy is far lower than that of the proposed algorithms, as easily seen by the naked eye and further witnessed by the corresponding plots of the relative error, with some tendency towards jittery and erratic behaviour. The adaptive histogram based algorithm of Schmeiser and Deutsch~\cite{SchmDeut1977} performed comparatively well on streams~2 and~3. On this account it may be surprising to observe its complete failure at producing a meaningful estimate in the case of stream~1. In fact the behaviour the algorithm exhibited on this data set is most useful in understanding the algorithm's failures modes. Notice at what points in time the estimate would shoot widely. After inspecting the input data it is readily observed that in each case this behaviour coincides with the arrival of a datum which is much larger than any of the historical data (and thus the range of the histogram). What happens then is that in re-scaling the histogram by such a large factor, many of the existing bins get `squeezed' into only a single bin of the new histogram, resulting in a major loss of information. When this behaviour is contrasted with the performance of the algorithms we proposed in this paper, the importance of the maximal entropy principle as the foundational idea is easily appreciated; although our algorithms too readjust their bins upon the arrival of each new datum, the design of our histograms ensures that no major loss of information occurs regardless of the value of new data.

\begin{figure}[htb]
  \centering
  \subfigure[12 bins]{\includegraphics[trim=0cm 0cm 22cm 0cm, clip=true, width=0.18\textwidth]{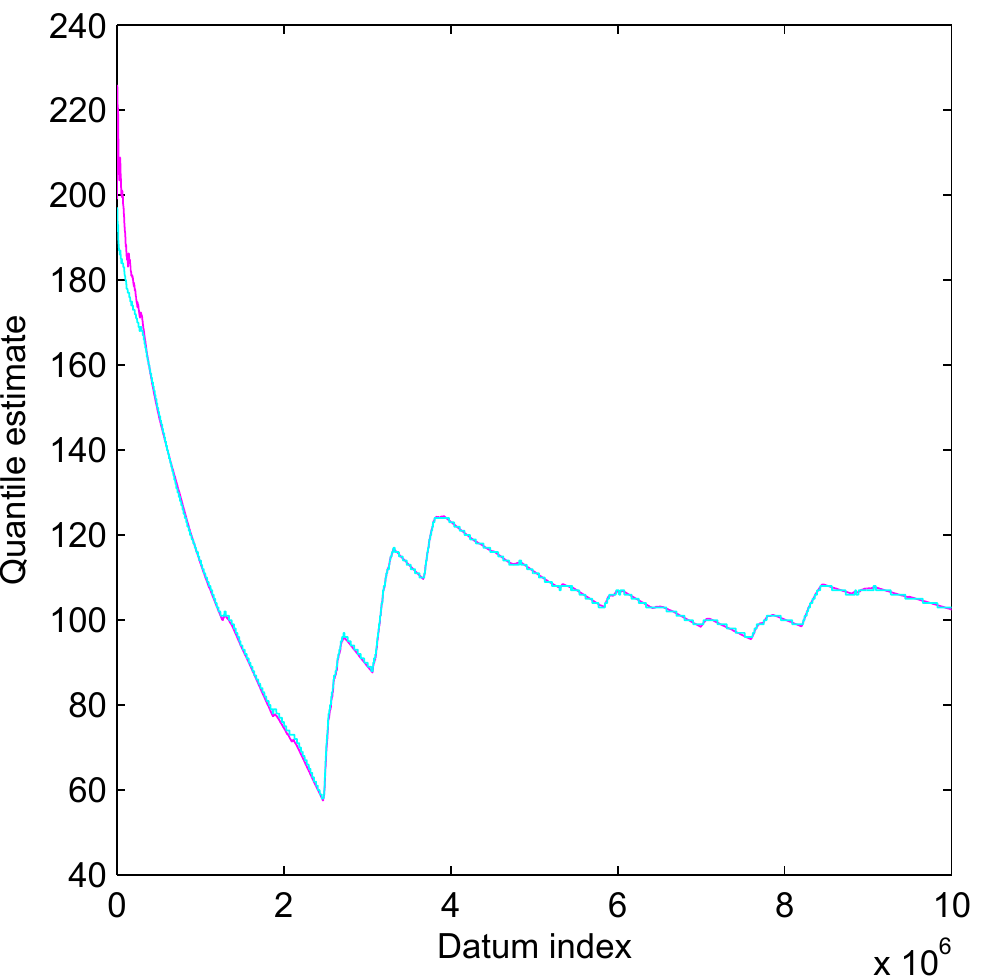}}
  \subfigure[25 bins]{\includegraphics[trim=0cm 0cm 22cm 0cm, clip=true, width=0.18\textwidth]{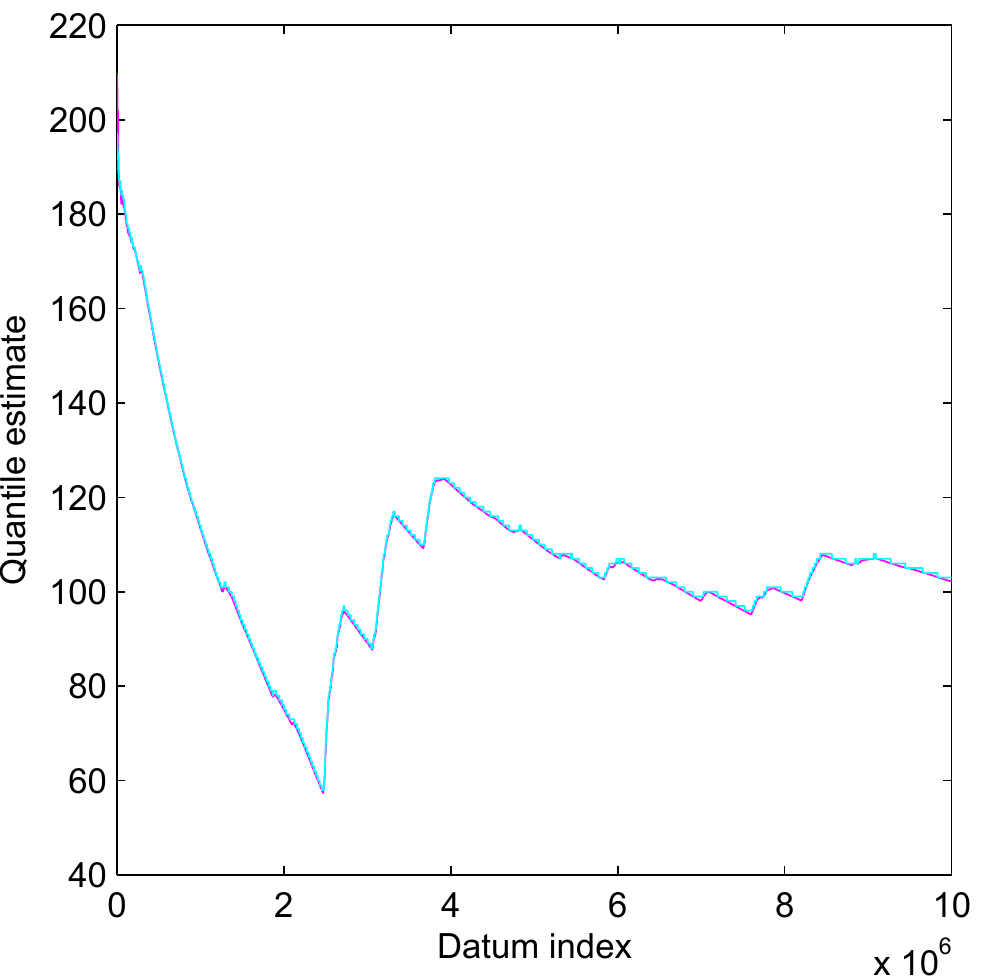}}
  \subfigure[50 bins]{\includegraphics[trim=0cm 0cm 22cm 0cm, clip=true, width=0.18\textwidth]{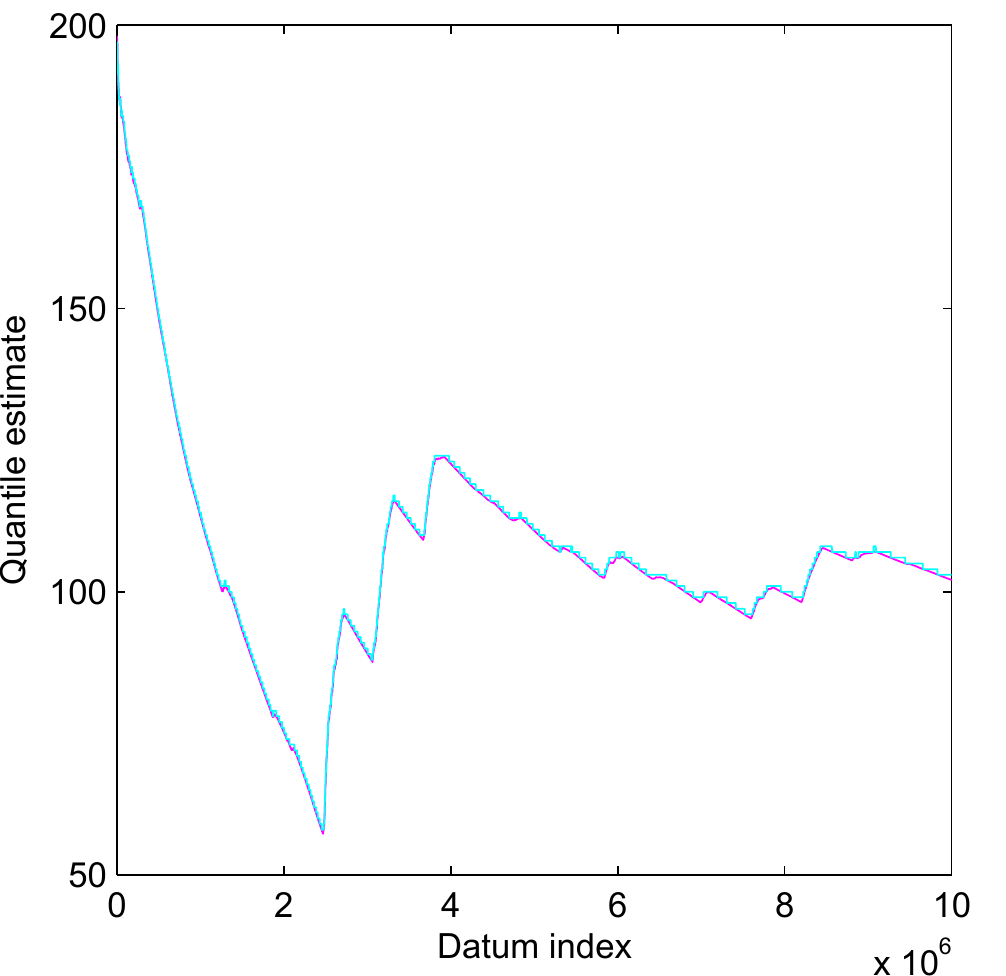}}
  \subfigure[100 bins]{\includegraphics[trim=0cm 0cm 22cm 0cm, clip=true, width=0.18\textwidth]{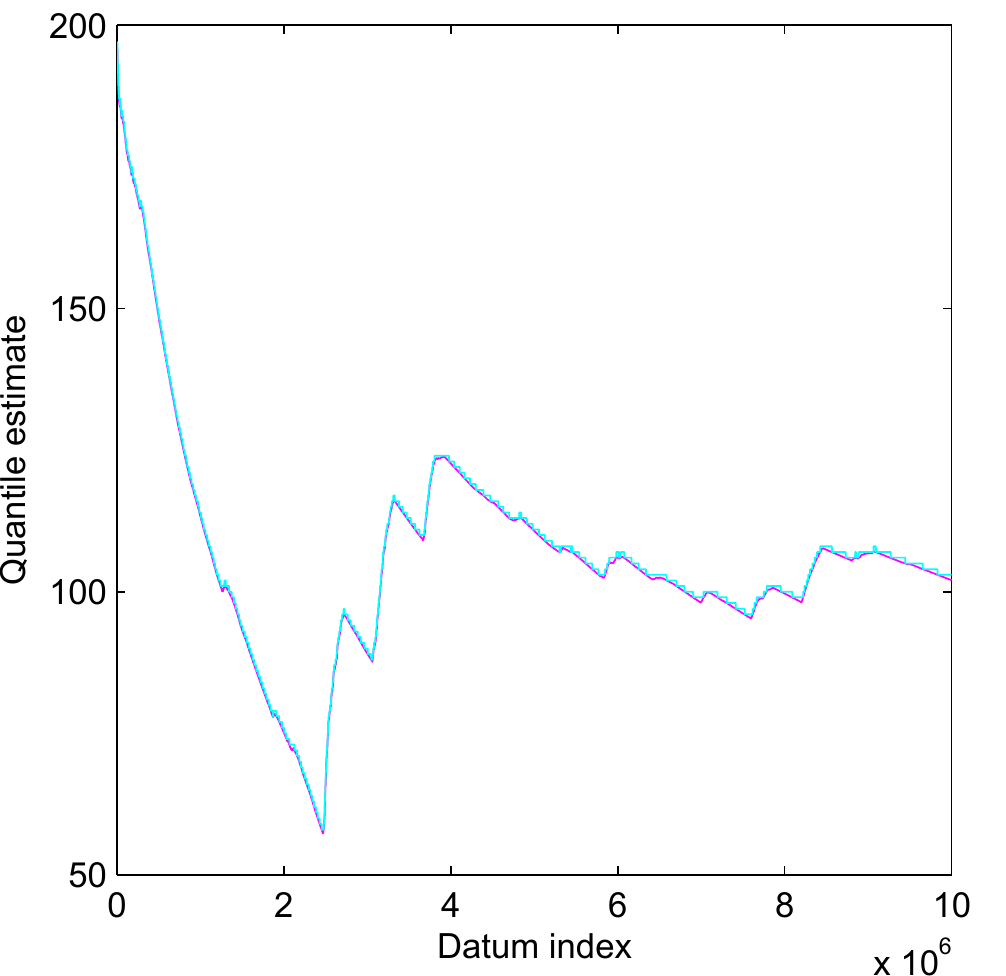}}
  \subfigure[500 bins]{\includegraphics[trim=0cm 0cm 22cm 0cm, clip=true, width=0.18\textwidth]{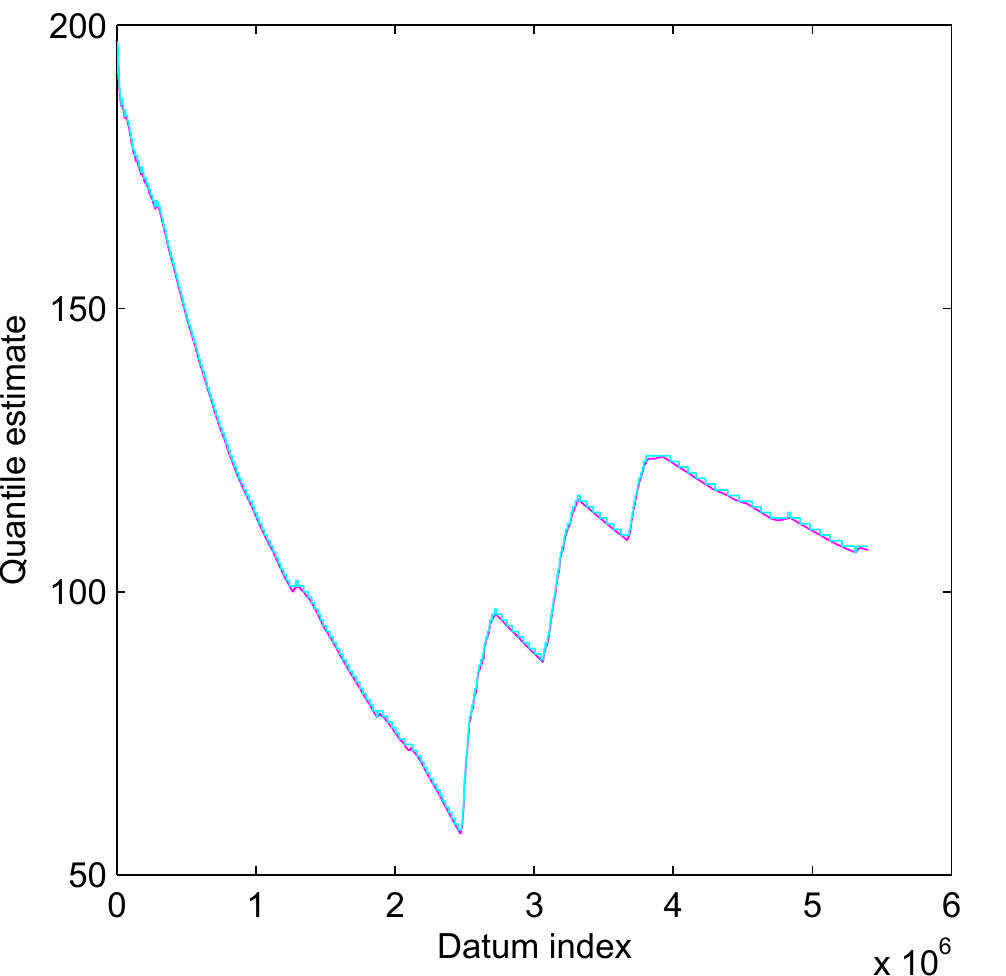}}
  \caption{ Running estimate (purple) of the 0.99-quantile on stream~2 produced using our data-aligned adaptive histogram algorithm (ground truth is shown in cyan). }
  \label{f:reduction}
  \vspace{-0pt}
\end{figure}

Considering the outstanding performance of our algorithms, and in particular the data-aligned histogram-based approach, we next sought to examine how this performance is affected by a gradual reduction of the working memory size. To make the task more challenging we sought to estimate the 0.99-quantile on the largest of our three data sets (stream~2). Our results are summarized in Table~\ref{t:res990}. This table shows the variation in the mean relative error as well as the largest absolute error of the quantile estimate for the proposed data-aligned histogram-based algorithm as the number of available bins is gradually decreased from 500 to 12. For all other methods, the reported result is for $n=500$ bins. It is remarkable to observe that the mean relative error of our algorithm does not decrease at all. The largest absolute error does increase, only a small amount as the number of bins is reduced from 500 to 50, and more substantially thereafter. This shows that our algorithm overall still produces excellent estimates with occasional and transient difficulties when there is a rapid change in the quantile value. Plots in Fig.~\ref{f:reduction} corroborate this observation.

\begin{table}
  \vspace{-5pt}
  \small
  \centering
  \renewcommand{\arraystretch}{1}
  \caption{ Summary of experimental results for the estimation of 0.99-quantile on stream~2. }
  \vspace{-5pt}
  \begin{tabular}{lc|cc}
  \Hline
  \multicolumn{2}{c|}{Method} & Mean relative error & Absolute $L_\infty$ error\\
  \hline
   \multirow{5}{*}{\rotatebox{90}{Proposed}}
   \multirow{5}{*}{\rotatebox{90}{data-aligned}}
   \multirow{5}{*}{\rotatebox{90}{bins w/ bin no.}}
                                                             & 500 & 0.5\% & 2.43 \\
                                                             & 100 & 0.5\% & 2.45 \\
                                                             &  50 & 0.5\% & 3.01 \\
                                                             &  25 & 0.4\% & 14.48 \\
                                                             &  12 & 0.5\% & 28.83 \\
   \hline
    $P^2$ algorithm \cite{JainChla1985} && 45.6\% & 112.61 \\
   \hline
    Random sample \cite{Vitt1985}       && 17.5\% & 64.00 \\
    \hline
    Equispaced bins \cite{SchmDeut1977} && 0.9\%  & 76.88 \\
  \Hline
  \end{tabular}
  \label{t:res990}
  \vspace{-10pt}
\end{table}

\section{Summary and conclusions}
We introduced two novel algorithms for the estimation of a quantile of a data stream when the available working memory is limited. The proposed algorithms were evaluated and compared against the existing alternatives described in the literature using three large data streams. The highly non-stationary nature of our data was shown to cause major problems to the existing algorithms, often leading to grossly inaccurate quantile estimates; in contrast, our methods were virtually unaffected by it. Our experiments demonstrate that the superior performance of our algorithms can be maintained effectively while drastically reducing the working memory size in comparison with the methods from the literature.

\bibliographystyle{unsrt}
\bibliography{./my_bibliography}
\end{document}